\documentclass[twocolumn,showpacs,pra,aps, longbibliography]{revtex4-2}
\usepackage{amsfonts}
\usepackage{amsmath}
\usepackage{physics}
\usepackage{graphicx}
\usepackage[unicode]{hyperref}
\hypersetup{
   unicode=true,          %
   plainpages=false,
   colorlinks=true,       
   citecolor=blue,        
   urlcolor=blue
}
\usepackage[caption=false,position=top,singlelinecheck=off,justification=raggedright]{subfig}
\usepackage{color}

\begin{document}

\title{Discrete time crystals and classical logic gates in coupled open Dicke models}

\author{Emmanuel D.G. U}
\email[]{edu@up.edu.ph}
\affiliation{National Institute of Physics, University of the Philippines, Diliman, Quezon City 1101, Philippines}

\author{Roy D. Jara Jr.}
\affiliation{National Institute of Physics, University of the Philippines, Diliman, Quezon City 1101, Philippines}

\author{Jayson G. Cosme}
\email[]{jcosme@nip.upd.edu.ph}
\affiliation{National Institute of Physics, University of the Philippines, Diliman, Quezon City 1101, Philippines}

\date{\today}

\begin{abstract}

We investigate the dynamics of coupled discrete time crystals (DTCs) in the coupled Dicke model or Dicke dimer (DD) with dissipation. By numerically simulating the mean-field dynamics of the system, we show that coupling two DTCs decreases their oscillation amplitude following a transient relaxation period. For sufficiently strong coupling, the coupled DTCs become unstable, resulting in the system relaxing towards the normal phase (NP). We demonstrate that the relaxation time diverges near the critical coupling strength, suggesting a continuous phase transition. Using the truncated Wigner approximation (TWA), we explore how the dynamics are affected by the introduction of quantum fluctuations. We also consider a scenario in which an NP is briefly coupled to a DTC, seeding the NP to transform into a DTC that is either synchronized or anti-synchronized relative to the DTC. We investigate the robustness of this phenomenon in the presence of noise. Using these results, we propose a scheme to perform NAND and NOR operations with three DTCs by sequentially coupling the input and output sites. We identify the parameter regimes in which the logic gate architecture remains valid, and investigate how the protocol is affected by the presence of quantum fluctuations.

\end{abstract}

\maketitle

\section{Introduction}

Developments in cold-atom experiments have allowed for the realization of exotic non-equilibrium phases of matter, an example of which is the DTC \cite{else_2020_dtcreview1, zaletel_2023_dtcreview2, Sacha_2017_dtcreview}. This dynamical phase emerges from the spontaneous breaking of discrete time translation symmetry imposed by a periodic drive, resulting in a robust subharmonic oscillation of the system's order parameter, with the period-doubling DTC being a fundamental example \cite{yao_2017_dtcs, else_2020_dtcreview1, zaletel_2023_dtcreview2, else_2016_dtcs, lazarides_2020_theodtc}. DTCs have been theoretically and experimentally investigated in many different platforms \cite{else_2020_dtcreview1, zaletel_2023_dtcreview2, Sacha_2017_dtcreview}, such as, but not limited to, superconductors \cite{jayson_2020_superconductingdtc, collado_2021_superconductingdtc, collado_2023_superconductingdtc}, spin systems \cite{sullivan_2020_expdtc, zhang_2017_experimentaldtc1, gambetta_2019_theodtc, khemani_2016_dtcs, chan_2015_spindtc, frey_2022_spindtc, owen_2018_spindtc, krishna_2023_spindtc, munozarias_2022_spindtc, russomanno_2017_spindtc, keyserlingk_2016_spindtc, frey_2022_superconductingdtcs, euler_2024_spindtcs, switzer_2025_spindtcs, nurwantoro_2019_spindtcs}, bosonic systems \cite{sacha_2015_dtcs, nie_2023_dickedtc, hans_2021_dickedtc, zhu_2019_dickedtcs, gong_2018_theoreticaldtc, yang_2021_bosondtc, bakker_2022_bosondtc, jayson_2019_bosondtc, richelle_2022_bosondtc, chitra_2015_bosondtcdicke, Jager_2024_bosondtcdickedtc, skulte_2021_bosondtcthreeleveldicke, kongkhambut_2021_bosondtcthreeleveldicke, rovny_2018_experimentaldtc2}, and networks of classical oscillators \cite{yao_2020_classicaldtcs, heugel_2019_classicaldtcs, nicolaou_2021_classicaldtc, heugel_2023_classicaldtc2, yithomas_2024_classicaldtc}. While interactions of multiple continuous time crystals have been explored \cite{sai_2022_ctcinteraction, Paulino_2026_ctcinteraction, greilich_2025_ctcinteraction, Postavova_2026_ctcinteraction}, the fundamental situation of coupling dissipative DTCs in the Dicke model remains largely unexplored.

It has been shown in Refs.~\cite{roy_2024_parametricmapping, roy_2025_kzmdtc} that the period-doubled DTCs in fully-connected open spin-boson systems in the mean-field or thermodynamic limit can be understood as a parametric resonance, in which the system can be effectively mapped onto a resonantly driven dissipative parametric oscillator. This mapping opens the question of whether the two degenerate DTC states of these periodically-driven systems can be used as logical bits, akin to the classical bits emulated in a dissipative parametric oscillator \cite{leuch_2016_posymmbreaking, fabiani_2022_pobits, mahboob_2008_pobits, nosaneichler_2019_pobits, frimmer_2019_notpoflipping, elyasi_2022_pobits, thegoto_1959_pobits}. While it has been demonstrated that DTCs emulating a single classical bit can be manipulated to perform the simplest bit operation, a bit-flip \cite{roy_2025_atp}, it remains to be seen whether more complex bit operations can be performed in coupled DTCs, similar to those performed using atomtronic setups \cite{ludwig_2025_paper1, ludwig_2025_paper2}.

\begin{figure}
    \centering
    \includegraphics[scale = 0.29]{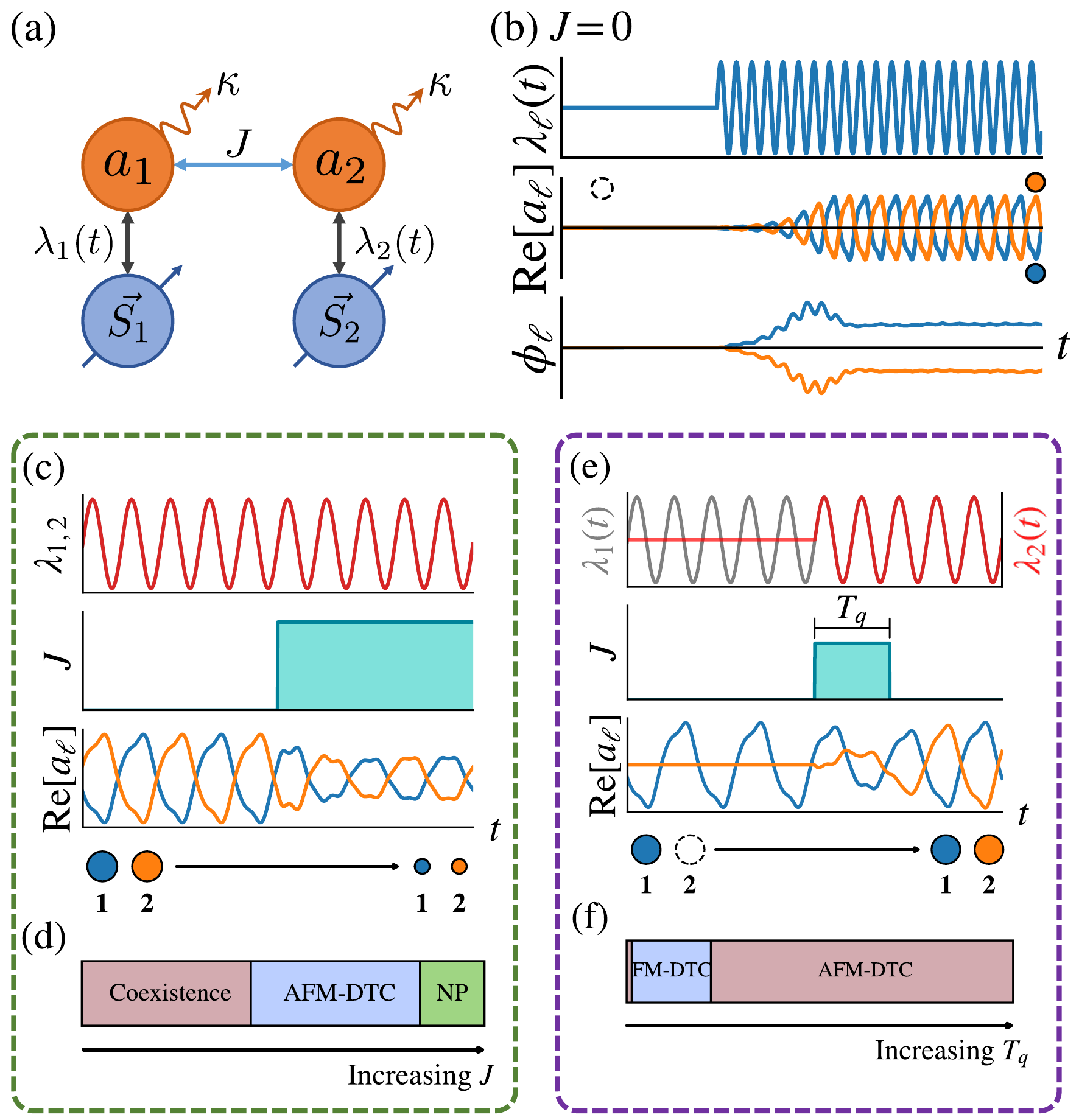}
    \caption{(a) Sketch of the DD. (b) (Top panel) The light-matter coupling $\lambda_i(t)$. (Middle panel) The real part of the bosonic field $a(t)$ for the two symmetry broken DTC states of the Dicke model, represented by the blue (dark) and orange (light) circles. The hollow circles represents the NP. (Bottom panel) The absolute time phases for the DTC states. (c) and (e) (Top panel) The light-matter coupling $\lambda_i(t)$ for each site. (Middle panel) The coupling strength $J$ between two sites in the DLM. (Bottom panel) Exemplary dynamics for $\mathrm{Re}[a_\ell]$ following the protocol shown in the panels above. The relative sizes of the circles represent the oscillation amplitude of the DTCs. Sketch of the phase diagram for varying (d) $J$ and (f) $T_q$}
    \label{fig:fig1}
\end{figure}

In this work, we explore the dynamics of coupled DTCs forming in a dimer of fully-connected spin-boson systems. To this end, we consider the periodically driven open DD, which describes the behavior of two bosonic systems coupled to an ensemble of two-level systems, or qubits, as shown in Fig.~\ref{fig:fig1}(a) \cite{zou_2014_dlm}. The two degenerate DTC states can be accessed by periodically driving the light-matter interaction strength $\lambda$, as shown in Fig.~\ref{fig:fig1}(b). For the protocol shown in Fig.~\ref{fig:fig1}(c), we find that coupling the two sites modifies the oscillation amplitudes of the DTCs in each site depending on the initial state of the dimer and the dissipation strength $\kappa$. Here, we initialize our system either as a ferromagnetic (FM) DTC, wherein both sites are oscillating in-phase, or an antiferromagnetic (AFM) DTC, wherein the sites are $\pi$-phase shifted from each other. A sketch of the phase diagram for fixed $\kappa$ and varying $J$ is depicted in Fig.~\ref{fig:fig1}(d), where we show transitions between FM-DTC, AFM-DTC, and NP. Furthermore, we demonstrate that these transitions are accompanied by a divergence in the relaxation time of the response of the system to the coupling, akin to a continuous phase transition. We then consider a pulse protocol for $J$, corresponding to a finite duration of coupling between a NP site and a site in the DTC phase, as illustrated in Fig.~\ref{fig:fig1}(e). This induces either synchronization or anti-synchronization with the DTC depending on the duration of the coupling $T_q$, as shown in  Fig.~\ref{fig:fig1}(f). Using these observations and components, we propose an architecture for the three-site DLM that recreates the universal logic operations NAND and NOR \cite{mano_2018_nanduniversal} by applying sequential pulses on the coupling between the three sites.

This paper is structured as follows. In Sec.~\ref{sec:sec2}, we introduce the system of interest and discuss the methods used to obtain the dynamics. Then, in Sec.~\ref{sec:sec3}, we consider the dynamics and stability of the coupled DTC states, as well as their behavior under the protocol shown in Fig.~\ref{fig:fig1}(c). In Sec.~\ref{sec:sec4}, we consider the synchronization and anti-synchronization of DTCs based on the protocol shown in Fig.~\ref{fig:fig1}(e), as well as its reliability in the presence of quantum fluctuations. Then, in Sec.~\ref{sec:sec5} we use these results to construct a protocol for the classical NAND logic gate. Lastly, we provide a conclusion and summary in Sec.~\ref{sec:sec6}.


\section{Theoretical Framework} \label{sec:sec2}
The DD models two dissipative bosonic systems coupled through the tunneling of bosons between the dissipative bosons across different sites. Each bosonic system interacts with an ensemble of $N$ two-level systems with a strength set by the spin-boson coupling strength $\lambda$. In the Markovian limit, the dynamics of the DD is described by the Lindblad master equation \cite{zou_2014_dlm}
\begin{equation}\label{eq:DickeMasterEq}
  \frac{\partial \hat{\rho}}{\partial t} = -i[\hat{H}/ \hbar, \hat{\rho}] + \frac{\kappa}{2} \sum_{\ell=1}^{M} (2 a_{\ell} \rho a_{\ell}^{\dagger} - \{ a_{\ell}^{\dagger}a_{\ell}, \rho \} ),
\end{equation}
where the DD considered here is a special case of the Dicke lattice model when $M = 2$ \cite{yilun_2024_dickedimer}.
\begin{align}\label{eq:DickeLatticeHamiltonian}
\begin{split}
    \frac{\hat{H}}{\hbar} &= \omega\sum_{\ell=1}^M \hat{a}_\ell^{\dagger} \hat{a}_\ell + \omega_0 \sum_{\ell=1}^M \hat{S}_\ell^z + \frac{2\lambda (t)}{\sqrt{N}} \sum_{\ell=1}^M (\hat{a}_\ell + \hat{a}^{\dagger}_\ell) \hat{S}_\ell^x \\
    &- \sum_{\ell=1}^MJ_\ell(\hat{a}^\dagger_\ell \hat{a}_{\ell+1} + \hat{a}_\ell \hat{a}_{\ell+1}^\dagger),
\end{split}
\end{align}
such that $\hat{H}_{DD} = \hat{H}_{}|_{M=2}$. The bosonic and spin excitation frequencies are denoted by $\omega$ and $\omega_{0}$, respectively. Here, the dissipation strength is $\kappa$, and the dissipative boson tunneling strength between the two sites is $J$. The operators $\hat{S}_\ell^{x,y,z}$ are the collective spin operators, whereas $\hat{a}_\ell$ and $\hat{a}^\dagger_\ell$ are the bosonic annihilation and creation operators for the $\ell$-th lattice site.

In the thermodynamic limit, the system's dynamics cam be accurately captured using the Holstein-Primakoff transformation, given by $\hat{S}_{\ell}^{z} = \hat{b}_{\ell}^{\dagger}\hat{b}_{\ell} - \frac{N}{2}$ and $\hat{S}_{\ell}^- = (\hat{S}_{\ell}^+)^\dagger = \sqrt{N} \sqrt{1 - \frac{\hat{b}^\dagger \hat{b}}{N}}\hat{b}$ \cite{holstein_1940_hpapprox, roy_2024_parametricmapping}. We use the approximation $S^{-}_{\ell} \approx 1 - \frac{1}{2}\frac{\hat{b}^\dagger \hat{b}}{N}$, which is valid for large $N$. This allows us to rewrite the spin operators in terms of bosonic creation and annihilation operators, leading to the following Hamiltonian:
\begin{align}\label{eq:HolsteinPrimakoffHamiltonian}
\begin{split}
  \frac{\hat{H}}{\hbar} &= \omega \sum_{\ell=1}^M \hat{a}_\ell^{\dagger} \hat{a}_\ell + \omega_0 \sum_{\ell=1}^M \hat{b}_\ell^{\dagger} \hat{b}_\ell - \sum_{\ell=1}^MJ_\ell(\hat{a}^\dagger_\ell \hat{a}_{\ell+1} + \hat{a}_\ell \hat{a}_{\ell+1}^\dagger) \\
  &+ \frac{2 \lambda(t)}{\sqrt{N}} \sum_{\ell=1}^M (\hat{a}_\ell + \hat{a}^{\dagger}_\ell) \left(\hat{b}_\ell + \hat{b}^{\dagger}_\ell - \frac{\hat{b}^{\dagger}_\ell(\hat{b}_\ell)^2}{2N} - \frac{(\hat{b}^{\dagger}_\ell)^\ell \hat{b}_\ell}{2N} \right).
\end{split}
\end{align}
To obtain the dynamics of the system, we employ TWA, a semiclassical approximation that maps Eq.~\eqref{eq:DickeMasterEq} to an effective Fokker-Planck equation describing the approximate quantum dynamics of the system in phase space \cite{polkovnikov_2010_twa}. This approximation, which is valid for finite $N \gg 1$, yields the following Heisenberg-Langevin equations of motion for the bosonic fields \cite{tuquero_2024_twaref, emman_2026_singlepulse}:
\begin{align}
    i \frac{da_\ell}{dt} &= \omega a_\ell + \lambda(b_\ell + b_\ell^*) \left(1 - \frac{|b_\ell|^2}{2N} \right) - J(a_{\ell + 1} + a_{\ell -1}) \notag \\
    &\quad -i\kappa a_\ell + \xi_\ell (t),\label{eq:motion_a} \\
    i \frac{db_\ell}{dt} &= \omega_0 b_\ell + \lambda(a_\ell + a_\ell^*) \left( 1 - \frac{1}{2N} (2|b_\ell|^2 + b_\ell^2)\right), \label{eq:motion_b}
\end{align}
where $a_{\ell}$ and $b_{\ell}$ are the complex variables associated with the operators $\hat{a}_{\ell}$ and $\hat{b}_{\ell}$, respectively. First-order quantum correlations manifest in the equations of motion through the stochastic noise term $\xi(t)$, which satisfies $\langle \xi_{\ell}\rangle=0$ and $\left\langle \xi^* (t) \xi (t')\right\rangle = \kappa \delta_{\ell, \ell'} \delta(t - t')$ \cite{tuquero_2024_twaref}. Following TWA, we numerically integrate the coupled stochastic differential equations in Eq.~\eqref{eq:motion_a}-\eqref{eq:motion_b} using Heun’s method with an integration time step of $\Delta t = 0.07$. Taking the thermodynamic limit, $N \rightarrow \infty$, we recover the mean-field equations of motion by allowing $\xi(t) / \sqrt{N} \rightarrow 0$.

For static $\lambda = \lambda_0$ in the thermodynamic limit $N \rightarrow \infty$, the open Dicke model, with a Hamiltonian $\hat{H}_{DM} = \hat{H}|_{M = 1}$, hosts two equilibrium phases. For $\lambda < \lambda_c = \frac{1}{2}\sqrt{\frac{\omega_0}{\omega} (\kappa^2 + \omega^2)}$, the system remains in the NP, which is characterized by the absence of macroscopic occupations of the bosonic mode $a_{\ell}$, and a fully polarized collective spin $\langle S_z \rangle$ in the -z direction, corresponding to $b_\ell = 0$. For $\lambda > \lambda_c$, the system transitions into the superradiant phase, in which both $a_{\ell}$ and $b_{\ell}$ exhibit macroscopic occupation. When $\lambda$ is periodically driven with the form $\lambda(t) = \lambda_{0}(1 + A\sin(\omega_{d} t))$, the open Dicke model can also host two symmetry broken states associated with a period-doubled DTC \cite{gong_2018_theoreticaldtc, nie_2023_dickedtc, hans_2021_dickedtc, zhu_2019_dickedtcs, chitra_2015_bosondtcdicke, skulte_2024_experimentaldickedtc, Jager_2024_bosondtcdickedtc}, as shown in Fig.~\ref{fig:fig1}(b). The period-doubling response manifests on both the dynamics of $a_{\ell}$ and $b_{\ell}$ \cite{roy_2024_parametricmapping}. 

We track the dynamics of the DTC by considering the dynamics of $a_{\ell}$. In a DTC state, the dynamics of $a_{\ell}$ can be expressed as a modulated waveform,
\begin{align}\label{eq:modulated_waveform}
a_\ell(t) &= r_\ell(t) \cos{\left(\omega_R t + \phi_\ell (t) \right)},
\end{align}
where $\omega_R = \omega_d / 2$ is the response frequency of the DTC, $r_{\ell}$ is a slowly varying amplitude, and $\phi_{\ell}(t)$ is the absolute-time phase \cite{roy_2025_atp, hoch_2021_atp, simula_2024_atp, catalini_2022_atp}. As shown in Fig.~\ref{fig:fig1}(b), when the system is in a DTC, $\phi_{\ell}$ acquires a constant value that is chosen from its two possible values that are $\pi$ shifted from one another. These two values correspond to the two degenerate DTC states of the system, and thus, we can use $\phi_{\ell}$ to distinguish between them. We can extract $\phi_{\ell}$ by performing a demodulation, in which we perform a moving weighted average over one driving cycle, yielding
\begin{align}
\frac{\omega_R}{\pi} \int_{t}^{t+\frac{2\pi}{\omega_R}} a_\ell(\tau) &e^{-i\omega_R\tau}\,d\tau = \frac{\omega_R}{2\pi} \int_{t}^{t+\frac{2\pi}{\omega_R}} r_\ell(\tau)e^{i\phi_\ell (\tau)}\,d\tau \nonumber\\
&\quad+ \frac{\omega_R}{2\pi} \int_{t}^{t+\frac{2\pi}{\omega_R}} r_\ell(\tau)e^{ -i(2\omega_R\tau+\phi_\ell (\tau))}\,d\tau,
\end{align}
wherein $e^{-i\omega_{R}\tau}$ serves as the weights in the moving average. Provided that $r_\ell (t)$ and $\phi_\ell (t)$ vary slowly compared to the carrier period, the first term on the right-hand side is approximately constant over the averaging period, and the integral simply returns the integrand scaled by a factor of $2\pi / \omega_R$, which cancels exactly with the prefactor outside the integral. The second term averages to $0$ due to its rapid oscillations at frequency $2\omega_R$. Thus, we can define the quantity
\begin{align}\label{eq:atp_fw}
    a_{\ell,  R} (t) \equiv \frac{\omega_R}{\pi} \int_t^{t+\frac{2\pi}{\omega_R}} e^{-i\omega_R \tau} a_\ell(\tau) d\tau \approx r_\ell(t) e^{i \phi_\ell(t)} ,
\end{align}
where $a_{\ell, R}$ denotes the complex amplitude associated with the period-doubled signal at the subharmonic frequency $\omega_R = \frac{1}{2}\omega_d$. The absolute-time phase is then the complex phase of $a_{\ell, R}$, $\phi_{\ell} = \mathrm{arg}[a_{\ell, R}]$.

\section{Dynamics of coupled discrete time crystals}\label{sec:sec3}

\begin{figure}[t]
    \centering
    \includegraphics[scale = 0.28]{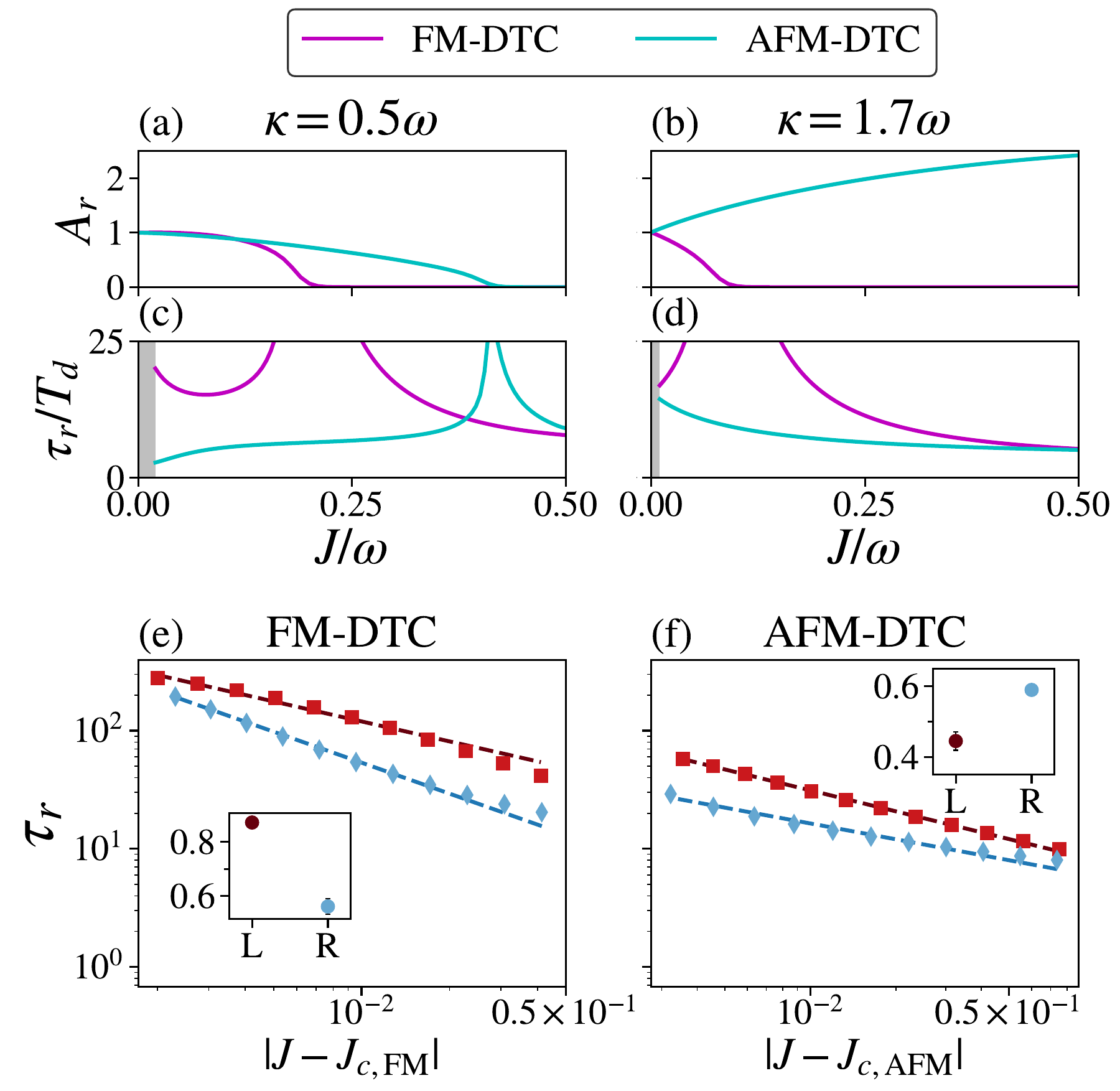}
    \caption{(a), (b) DTC response amplitude, $A_{r}$, and (c), (d)relaxation time of the response amplitude, $\tau_{r}$, after coupling the sites. (e), (f) Power law-scaling of $\tau_r$ for the FM-DTC and AFM-DTC near the critical points $J_{c, \mathrm{FM}}$ and $J_{c, \mathrm{AFM}}$, respectively, for $\kappa = 0.5\omega$. The square (diamond) datapoints are taken from the right (left) side of $J_{c, \mathrm{FM/AFM}}$. The inset plots show the critical exponents obtained from the power-law fit. The other parameters used are $\{\omega_0, \omega_d, A, N, \lambda_0\} = \{\omega, 0.8\omega, 0.5, 10^7, 0.85\lambda_c\} $.}
    \label{fig:fig2}
\end{figure}

\subsection{Continuous coupling}\label{sec:sec3a}

We begin our analysis by considering the dynamics of two DTCs coupled after their respective absolute-time phase has reached a steady state, as shown in Fig.~\ref{fig:fig1}(c). The mean-field dynamics suggest that right after the coupling, the response amplitude of the system undergoes a transient response for some time $\tau_r$ before eventually approaching a steady-state $\mathrm{Re}[a_{\ell, R}(t_f)]$ that differs from the initial response amplitude $\mathrm{Re}[a_{\ell, R}(t_i)]$. Here, $t_f$ denotes the final simulation time, chosen so that the system is able to attain a steady state after being coupled, and $t_i$ denotes a time when the system is in a steady state prior to the coupling. We seek to characterize this response of the system in terms of the relaxation time $\tau_r$ and the coupled steady-state amplitude $\mathrm{Re}[a_{\ell, R}(t_f)]$ across different parameter regimes. In the following, we characterize the state of the two-site DLM by the phase difference between the two sites $\Delta \phi \equiv |\phi_1 - \phi_2|$. We refer to the state with $\Delta \phi = 0$ as an FM-DTC, and that with $\Delta \phi = \pi$ as an AFM-DTC.

In Figs.~\ref{fig:fig2}(a) and \ref{fig:fig2}(b), we show the dependence of $A_r = |a_{\ell=1} (t_f)|/ |a_{\ell=1}(t_i)|$ on $J$ by solving the mean-field equations of motion for $\kappa = 0.5\omega$ and $\kappa = 1.7 \omega$, respectively, and classifying the dynamics of the coupled system according to the favored DTC configuration. For these values of $\kappa$, both AFM-DTC and FM-DTC initial states survive with a modified amplitude at small values of $J$. Increasing $J$ past the critical point $J_{c, \mathrm{FM}}$ pushes the FM-DTC into the NP while favoring the AFM-DTC, which remains in a stable DTC phase. For $\kappa = 0.5\omega$, there exists another critical point $J_{c, \mathrm{AFM}}$ at which the AFM-DTC becomes unstable and relaxes into the NP. Each of these transitions to the NP are accompanied by a divergence of the relaxation time $\tau_{r, \mathrm{FM / AFM}}$, as shown in Figs.~\ref{fig:fig2}(c)-(d), akin to divergences observed in continuous phase transitions in equilibrium. Here, we extract $\tau_{r, \mathrm{FM / AFM}}$ by fitting the response amplitude of the first site $r_1 (t)$ to the exponential form $f(t) = A_0 \exp(-t/\tau_{r, \mathrm{FM / AFM}}) + C$, where $A_0, C \in \mathbb{R}$. Motivated by these results, we fit the behavior of $\tau_{r, \mathrm{FM / AFM}}$ near the critical points $J_{c, \mathrm{FM / AFM}}$ to the power-law form $f(t) = D |J - J_{c, \mathrm{FM / AFM}}|^{-vz} + E$, where $D, E, vz \in \mathbb{R}$ \cite{roy_2025_kzmdtc, delcampo_2014_kzmreference}. In Fig.~\ref{fig:fig2}(e) and \ref{fig:fig2}(f), we display the lines of best fit as $J$ approaches $J_{c, \mathrm{FM / AFM}}$ from either the left $(-)$ or the right $(+)$ of the critical point when $\kappa = 0.5\omega$, along with the critical exponents $(vz)_{\pm, \mathrm{FM} / \mathrm{AFM}}$ found in the inset plots. We note that $(vz)_{-, \mathrm{FM/ AFM}} \neq (vz)_{+, \mathrm{FM/AFM}}$ and $(vz)_{\pm, \mathrm{FM}} \neq (vz)_{\pm, \mathrm{AFM}}$.

\begin{figure}
    \centering
    \includegraphics[scale = 0.29]{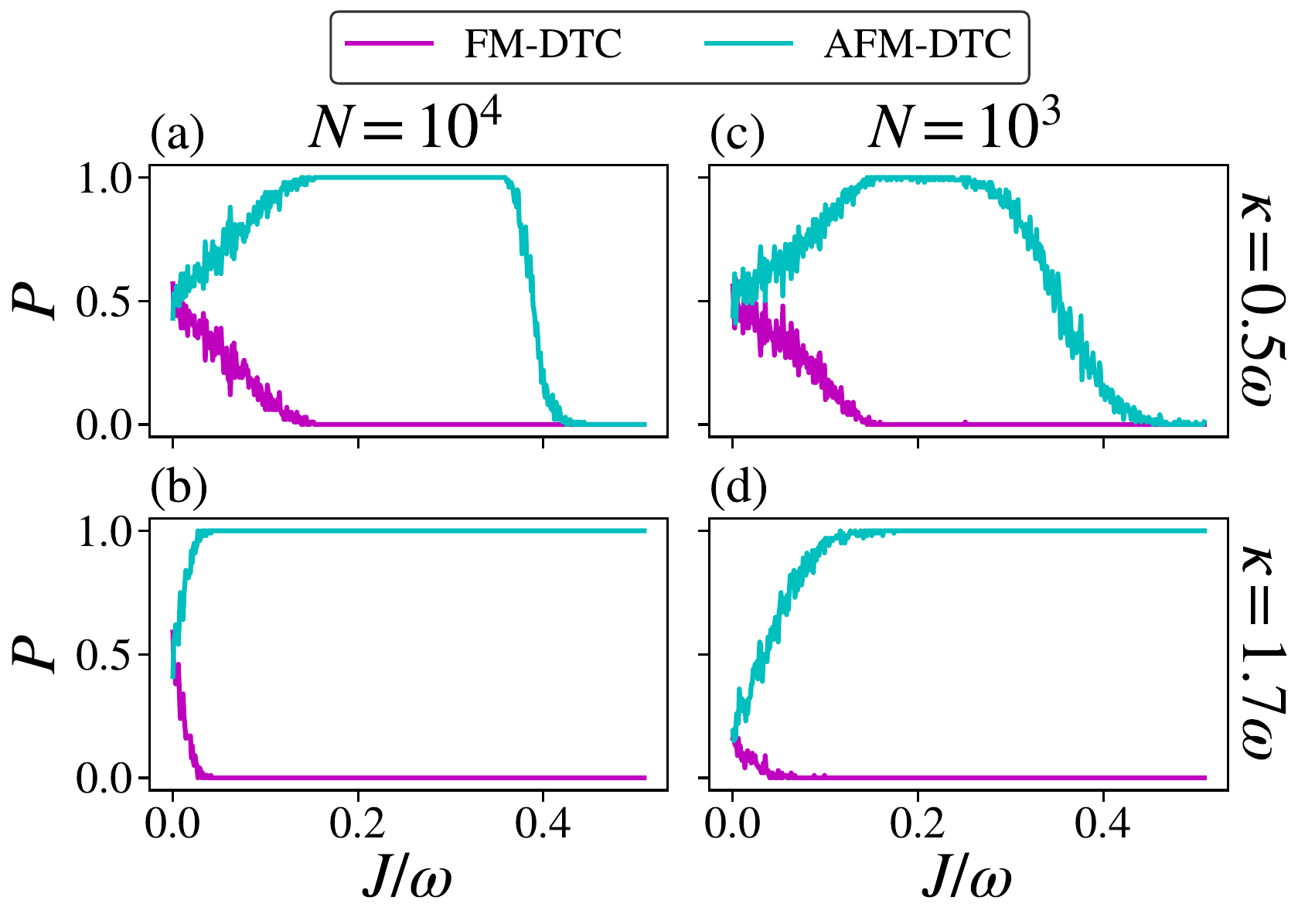}
    \caption{The probability $P$ of obtaining either an FM-DTC or AFM-DTC as the final steady state across 100 trajectories for (a), (c) $\kappa = 0.5\omega$ and (b), (d) $\kappa = 1.7\omega$. Here, the two sites are connected for the entire duration of the simulation. We consider a system size of $N = 10^4 $ for (a) and (b), while we set $N=10^3$ for (c) and (d). The other parameters are $\{\omega_0, \omega_d, A, N, \lambda_0\} = \{\omega, 0.8\omega, 0.5, 10^7, 0.85\lambda_c\} $.}
    \label{fig:fig3}
\end{figure}

In order to further probe the stability of the FM and AFM-DTC states, we investigate the dynamics of the system in the presence of quantum fluctuations by solving the coupled stochastic equations, Eqs.~\eqref{eq:motion_a} and \eqref{eq:motion_b}. Here, it is instructive to first consider the case when both sites are coupled from the start of the simulation. In Figs.~\ref{fig:fig3}(a) and \ref{fig:fig3}(b), we show the probability $P_{\mathrm{FM}/\mathrm{AFM}}$ of obtaining either an FM or AFM-DTC for the final steady state across 100 trajectories for $N = 10^4$. The bosonic modes are initialized in a vacuum, which can be numerically implemented by sampling the real and imaginary components of the complex random variable $O_{\ell} = (O_{\mathrm{R}, \ell} + iO_{\mathrm{I}, \ell}) / 2$ from a Gaussian distribution centered at zero satisfying $\langle O_{\ell, m}O_{\ell', m'}\rangle = \delta_{\ell, \ell}'\delta_{m, m'}$ for $O_{\ell} \in \{ a_{\ell}, b_{\ell}\}$ and $m \in \{ \mathrm{R}, \mathrm{I} \}$. For both $\kappa = 0.5\omega$ and $1.7\omega$, the probability of obtaining either state is approximately equal near $J = 0$, since the coupling is not strong enough to affect the final state of the system. As $J$ increases, the probability of obtaining an FM-DTC, $P_\mathrm{FM}$, gradually decreases until it reaches zero near $J_{c, \mathrm{FM}}$, indicating that the AFM-DTC becomes energetically preferred in this region. This is consistent with the findings shown in Fig.~\ref{fig:fig2}, since for $J > J_{c, \mathrm{FM}}$, the FM-DTC relaxes into the NP, in which the fluctuations eventually pushes the system towards the stable AFM-DTC states. Above $J_{c, \mathrm{AFM}}$ for $\kappa = 0.5\omega$, the AFM-DTC eventually relaxes towards the NP, leading to a vanishing $P_\mathrm{AFM}$. In Figs.~\ref{fig:fig3}(c) and \ref{fig:fig3}(d), we show the case when $N$ is decreased to $10^3$. Note that a decrease in the system size results in a decrease in the response amplitude of the period-doubled response in the bosonic fields $a_\ell (t)$, thereby increasing the relative strength of the fluctuations. For $\kappa = 0.5\omega$, shown in Fig.~\ref{fig:fig3}(c), we note that $P_\mathrm{FM}$ is not significantly affected by the reduction in $N$, whereas $P_\mathrm{AFM}$ starts decreasing at smaller values of $J$ compared to $N=10^4$. This is because as $J$ approaches $J_{c, \mathrm{AFM}}$, the oscillation amplitude of the AFM-DTC becomes more suppressed, allowing for the fluctuations to dominate the dynamics. When $N$ is decreased to $10^3$ for $\kappa = 1.7\omega$, shown in Fig.~\ref{fig:fig3}(d), the amplitude of the stable DTC states are even smaller due to the increased dissipation rate. Consequently, near $J = 0$, the fluctuations dominate, and the probability that a DTC survives in a single-noise realization is $P_{FM} + P_{AFM} \leq 1$. However, as seen in Fig.~\ref{fig:fig2}(b), increasing $J$ amplifies the amplitude of the AFM-DTC state and therefore increases its stability. As a result, $P_\mathrm{AFM}\rightarrow 1.0$ for large $J$ in Fig.~\ref{fig:fig3}(d).

\begin{figure}
    \centering
    \includegraphics[scale = 0.18]{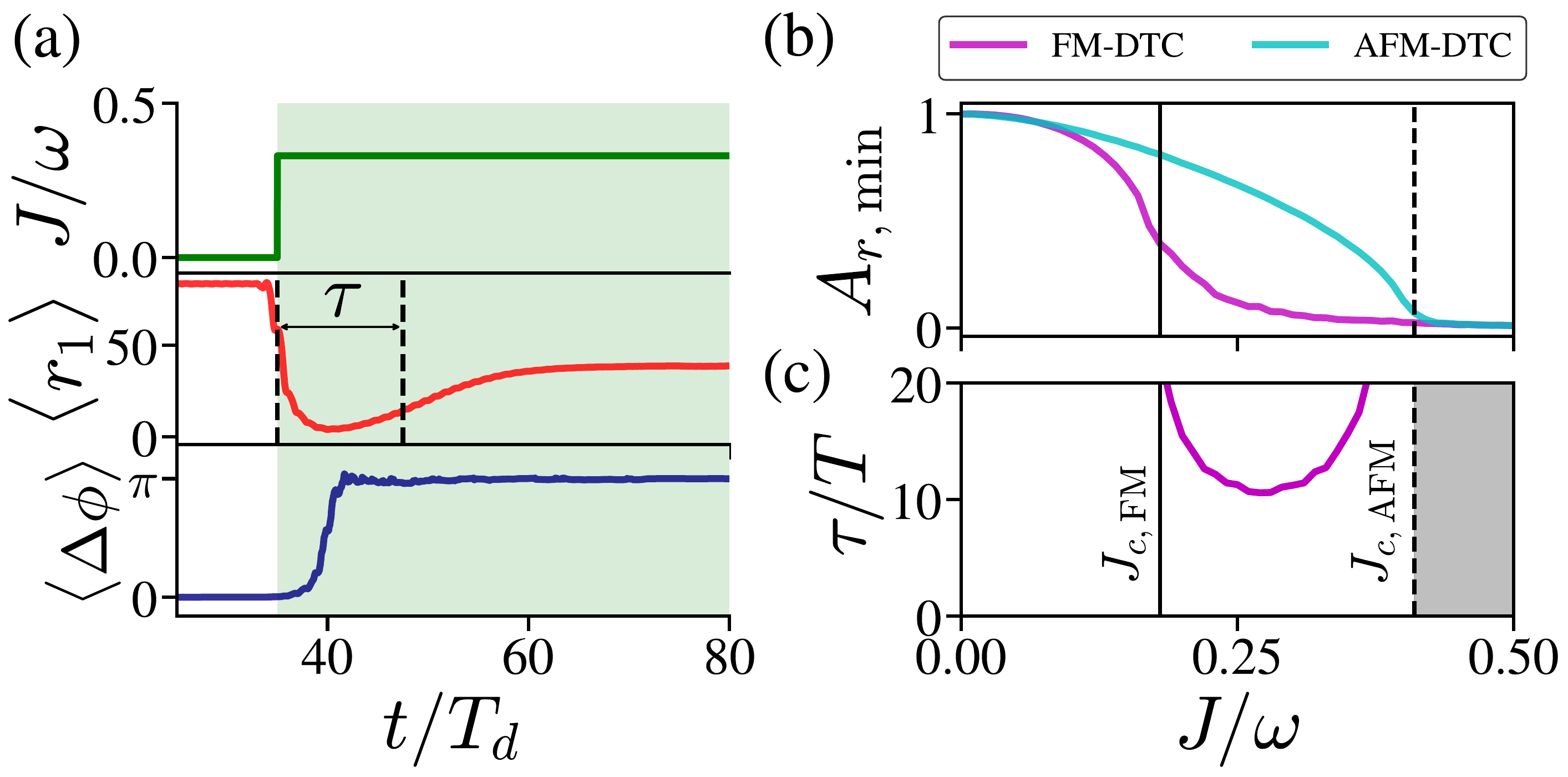}
    \caption{(a) (Top panel) The coupling strength $J$ between the two sites, (Middle panel) $\langle {r}_1 \rangle $, and (Bottom panel) $ \langle \Delta \phi \rangle$ as functions of time. The green-shaded region indicates when the two sites are coupled with strength $J = 0.33\omega$ (b) $A_{r, \mathrm{min}}$ and (c) the lifetime $\tau$ of the FM-DTC as a function of $J$ when $N = 10^5$. The other parameters used are $\{\omega_0, \omega_d, A, \lambda_0\, \kappa\} = \{\omega, 0.8\omega, 0.5, 0.85\lambda_c, 0.5\omega \} $.}
    \label{fig:fig4}
\end{figure}

We next examine how the presence of quantum fluctuations affects the continuous coupling protocol depicted in Fig.~\ref{fig:fig1}(c) for FM and AFM-DTC initial states. To this end, we suppress the noise term $\xi(t)$ in our simulations until an initial steady state is reached, otherwise the fluctuations can potentially affect the system as it propagates towards the desired initial state. This is equivalent to a post-selection process in which we select either an FM-DTC or AFM-DTC as the initial state before the protocol. In Fig.~\ref{fig:fig4}(a), we perform the protocol on the FM-DTC initial state and show the ensemble averages of the response amplitude of $a_{\ell = 1}$, $\langle r_{1} \rangle$, and the phase difference between the sites $\langle \Delta \phi \rangle = \langle \phi_{\ell = 1}(t) - \phi_{\ell = 2}(t) \rangle$ across 100 trajectories. Note that we have chosen $J$ such that $ J_{c, \mathrm{FM}} < J < J_{c, \mathrm{AFM}}$. The exemplary dynamics in Fig.~\ref{fig:fig4}(a) suggest that on average, the coupling initially pushes the system close to the NP. However, after some finite time $\tau$ has elapsed, the fluctuations push the system away from the NP, causing it to spontaneously choose between the two degenerate AFM-DTC states, for which $\langle \Delta \phi \rangle = \pi$. In order to quantify how closely the system approaches the NP, we show the quantity $A_{r, \mathrm{min}} = \min \langle {r}_1 \rangle / \max \langle {r}_1 \rangle$ in Fig.~\ref{fig:fig5}(b). For $J_{c, \mathrm{FM}} < J < J_{c, \mathrm{AFM}}$, we see that in contrast to the mean-field observations, the FM-DTC always hosts a non-zero amplitude, indicating that the fluctuations push the system into a stable AFM-DTC even before it can relax into the NP. For $J>J_{c, \mathrm{AFM}}$, $A_{r, \mathrm{min}}=0$ for both the FM-DTC and AFM-DTC, since only the NP is stable in this regime. Thus, the system always relaxes to the NP despite the presence of fluctuations. In Fig.~\ref{fig:fig4}(c), we show how the lifetime of the FM-DTC, $\tau$, varies as a function of $J$. For $J < J_{c, \mathrm{FM}}$, the FM-DTC remains stable with a modified amplitude, and so $\tau \rightarrow\infty$ in this region. When $J_{c, \mathrm{FM}} < J < J_{c, \mathrm{AFM}}$, the FM-DTC becomes unstable and acquires a finite lifetime, with the minimum $\tau$ being found near the midpoint of both critical points. This is because as $J$ approaches $J_{c, \mathrm{AFM}}$, the stability of the AFM-DTC diminishes, and so it takes longer for the FM-DTC to be pushed towards the AFM-DTC state. For $J > J_{c, \mathrm{AFM}}$, the FM-DTC initial state is no longer pushed into an AFM-DTC, and hence $\tau$ becomes undefined.

Given that we have identified how two initially uncoupled DTCs behave once coupled indefinitely, we next consider what happens when two initially uncoupled DTCs are coupled momentarily. This scenario will be essential for understanding the protocol for emulating a classical logic gate using DTCs.

\subsection{Pulsed coupling}\label{sec:sec4}

\begin{figure}
    \centering
    \includegraphics[scale = 0.27]{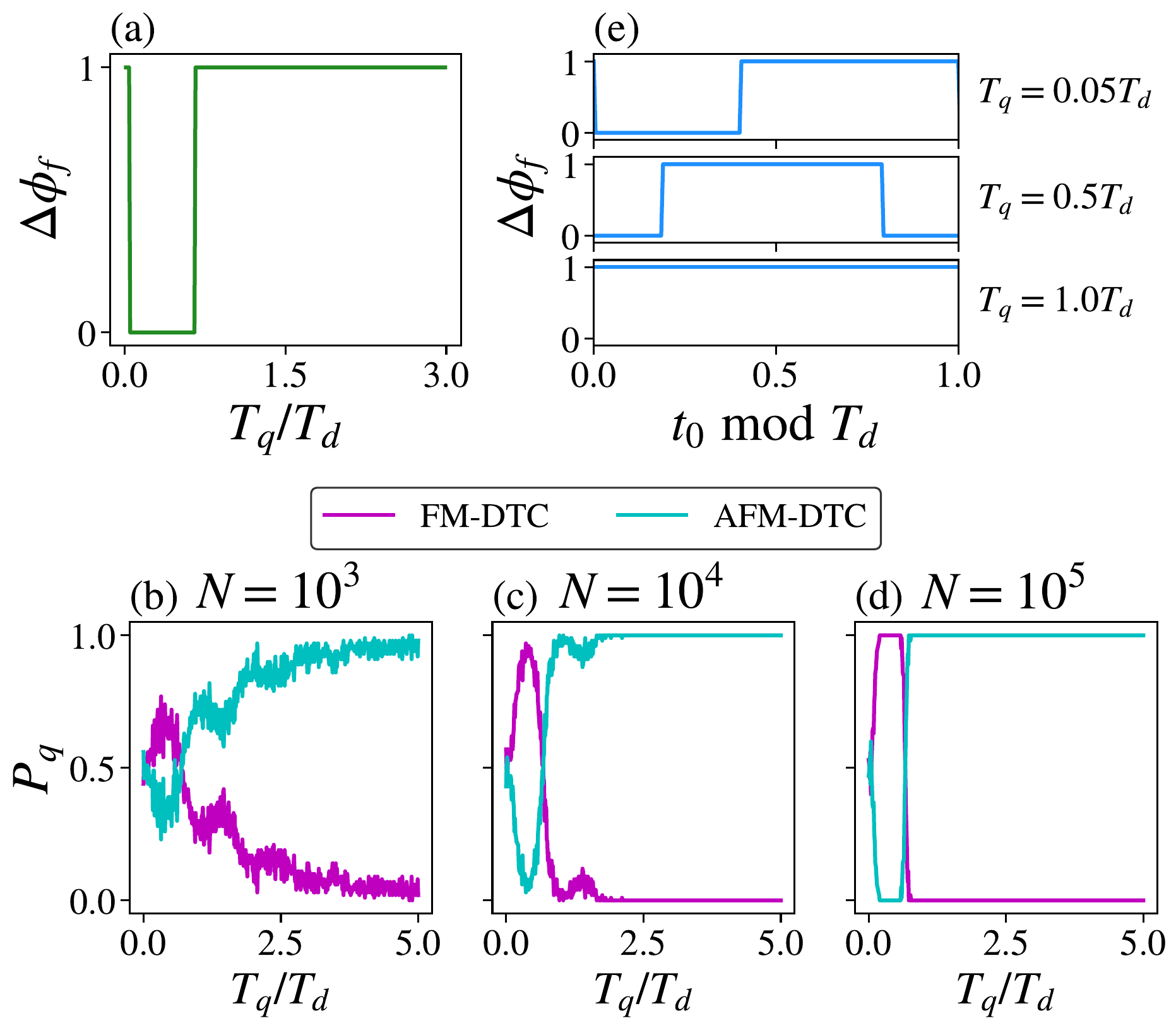}
    \caption{(a) Mean-field phase difference $\Delta \phi_{f}$ of the two sites at the final time as a function of $T_q$ for the pulsed coupling. (b) - (d) The probability $P$ of obtaining either an FM-DTC or an AFM-DTC when performing the protocol for (b) $N = 10^3$, (c) $N = 10^4$, and (d) $N = 10^5$. (e) Behavior of $\Delta \phi$ when the synchronization / anti-synchronization protocol is performed at different times in the driving cycle for (Top panel) $T_q = 0.05T_d$, (Middle panel) $T_q = 0.5T_d$, and (Bottom panel) $T_q = 1.0T_d$. The other parameters used are $\{\omega_0, \omega_d, A, N, \lambda_0\, \kappa\} = \{\omega, 0.8\omega, 0.5, 10^7, 0.85\lambda_c, 0.5\omega \} $.}
    \label{fig:fig5}
\end{figure}

We now consider a scenario in which the first site initialized in the NP is briefly coupled to another site in the DTC phase, as shown in Fig.~\ref{fig:fig1}(e). To this end, we remove the periodic driving in the first site while the system is uncoupled and set the spin-boson coupling to be less than $\lambda_{c}$ so that the system remains in NP. The exemplary dynamics in Fig.~\ref{fig:fig1}(e) demonstrate that if we simultaneously periodically drive the first site as we couple it to the other site for a finite duration $T_{q}$, the first site can be made to anti-synchronize with the pre-existing DTC, leading to the formation of an AFM-DTC. To identify the dependence with $T_q$ of the final DTC configuration after the pulsed coupling, we calculate the final phase difference between the DTCs in the two sites, $\Delta \phi_f \equiv |\phi_{\ell = 1}(t_f) - \phi_{\ell=2}(t_f) |$. In Fig.~\ref{fig:fig5}(a), we show how $\Delta \phi_f$ behaves as a function of $T_q$ for $\kappa = 0.5\omega$, and in the absence of quantum fluctuations. We identify a small region near $T_q = 0$ in which $\Delta \phi_f = 0$, signifying that the first site synchronizes with the DTC in the second site and creates an FM-DTC. Outside this region, we instead observe anti-synchronization, which results in an AFM-DTC with $\Delta \phi_f = \pi$.

We next examine how the synchronization and anti-synchronization is affected by the inclusion of quantum fluctuations. We again suppress the noise term $\xi(t)$ in our simulations until the protocol begins to mimic a post-selection of the initial DTC state in the second site, as explained in Sec.~\ref{sec:sec3a}, while we keep the first site in the NP until the pulsed coupling is applied. In Figs.~\ref{fig:fig5}(b) and \ref{fig:fig5}(c), we display the probabilities $P_{q, \mathrm{FM} / \mathrm{AFM}}$ that the protocol produces either an FM-DTC or AFM-DTC after the pulsed coupling as a function of $T_{q}$ for different $N$. Near $T_q = 0$, $P_{q, \mathrm{FM}}$ and $P_{q, \mathrm{AFM}}$ are approximately equal because the coupling duration is too short to sufficiently influence the final state of the system. Consequently, the quantum fluctuations seed the AFM-DTC and FM-DTC final states with approximately equal probability. This is in contrast to the mean-field regime, which predicts that $\Delta \phi_f = 0$ near $T_q = 0$, resulting in an AFM-DTC. As we increase $T_{q}$, the probability of obtaining an FM-DTC approaches a maximum value for intermediate $T_q$. For $N = 10^3$, shown in Fig.~\ref{fig:fig5}(b), the fluctuation strength is comparable to the oscillation amplitude of the DTCs, leading to a significant effect on the outcome of the protocol. Specifically, this causes $P_{q, \mathrm{FM}}<1$ even in the region wherein the mean-field solution predicts $\Delta \phi_f = 0$. As $T_q$ is increased further, the probability for the FM-DTC eventually vanishes, and the system is almost guaranteed to produce an AFM-DTC. In Fig.~\ref{fig:fig5}(c) and \ref{fig:fig5}(d), we show the effects of increasing $N$ to $N=10^4$ and $N= 10^5$, respectively. We find that for large $N$, the influence of quantum noise becomes weaker, causing the boundaries separating the regions with $\Delta \phi_f = 0$ and $\Delta \phi_f = \pi$ to approach the mean-field limit as they become sharper and more pronounced.

Previously, we have considered the case when the coupling is activated at the start of a driving cycle of $\lambda(t)$. We now examine how the initial time $t_0 = t\mod{T_d}$ of the coupling affects the result of the protocol. In Fig.~\ref{fig:fig5}(e), we show how $\Delta \phi_{f}$ depends on $t_0$ for various $T_q$ in the absence of quantum noise. For $T_q = 0.25T_d$, we see that $\Delta \phi_{f} = 0.0$ for small $t_0$. For large $t_0$, we instead observe a phase difference of $\Delta \phi_{f} = 1.0$. Finally, when we increase $T_{q}$ to $T_{q} = 1.0T_{d}$, we get $\Delta \phi_{f} = 1$ for all values of $t_0$, indicating that the dependence of the final state on $t_0$ is seen only for small $T_q$.

As we have demonstrated, we can readily manipulate the DTC configuration of the system by dynamically modifying the coupling between each site. We will now use these results to propose a logical gate architecture for DTCs in the next section.

\section{Classical Logic operations using coupled DTCs}\label{sec:sec5}

\begin{figure}\label{fig:Dicke_Validity}
    \centering
    \includegraphics[scale = 0.29]{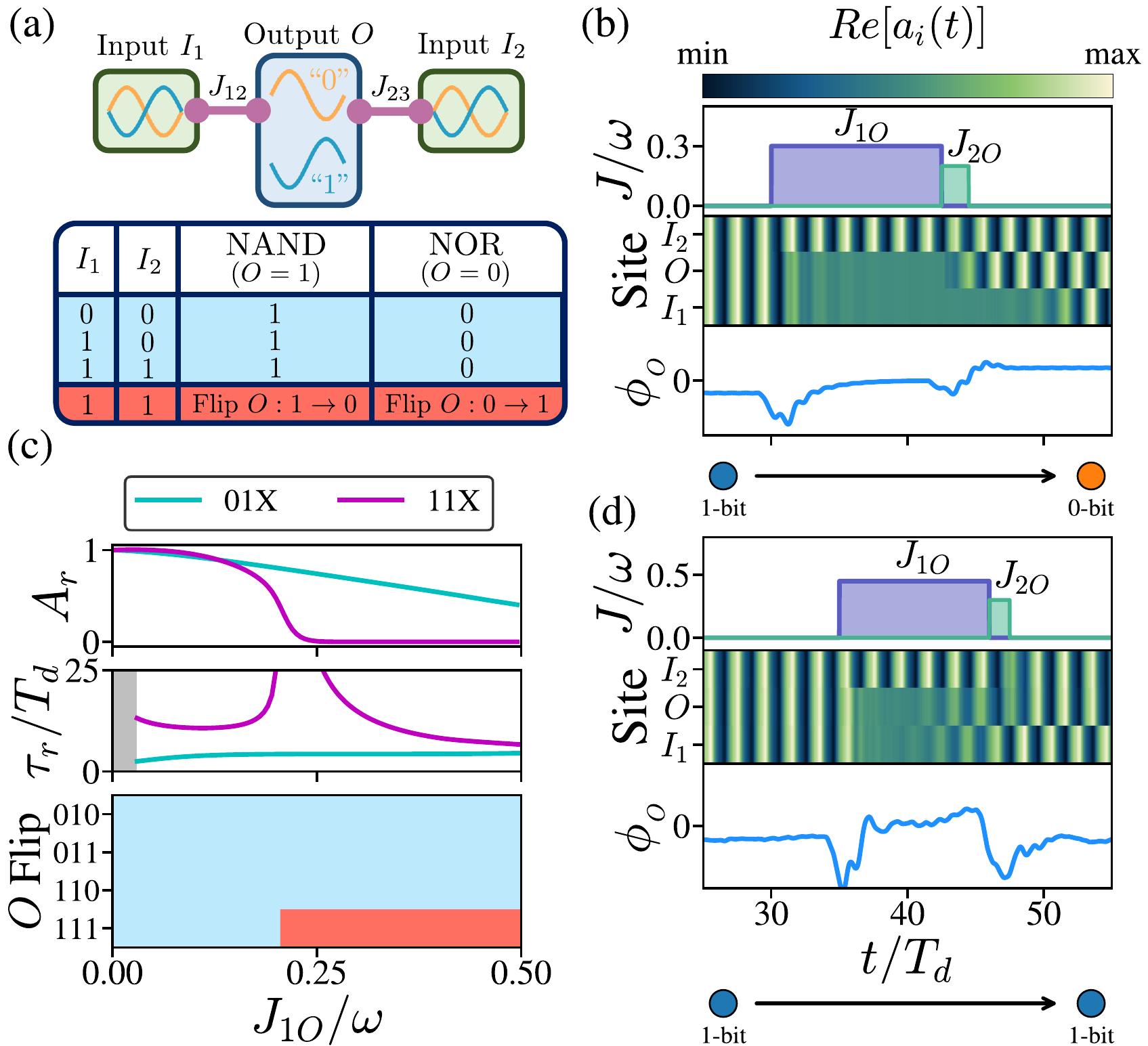}
    \caption{(a) Sketch of the proposed architecture for a NAND or NOR gate. (b) and (d) Exemplary dynamics showing the NAND gate operation for the input configuration [111]. (Top panel) The coupling $J_{1O}$ and $J_{2O}$, (Middle panel) the dynamics of $\mathrm{Re}[a_\ell(t)]$ for each site, and (Bottom panel) the absolute time phase $\phi_O$ for the output site. The colored markers underneath the bottom panel indicate the state of the output site. (b) Exemplary dynamics showing a successful NAND operation in the mean-field limit. (c) (Top panel) $A_r$, (Middle panel) $\tau_r$, and (Bottom panel) the flipping of the output site for each input configuration as functions of $J$, where the light-shaded (dark-shaded) region indicates a flip (no flip) from $1\rightarrow0$. (d) Exemplary dynamics depicting how the protocol can fail in the presence of quantum fluctuations when $N = 10^4$. The parameters used are $\{\omega_0, \omega_d, A, \lambda_0\} = \{\omega, 0.8\omega, 0.5, 0.9\lambda_c\} $.}
    \label{fig:fig6}
\end{figure}

Given the results in the previous section, we now propose a logic-gate architecture that implements a NAND or NOR operation between three spin-boson systems that are in a DTC phase. To highlight that the symmetry-broken states of the DTC can be treated as effective classical bits \cite{roy_2025_kzmdtc, roy_2025_atp}, we denote the state with $\phi < 0$ ($\phi > 1$) as the 1-bit (0-bit). Note that since these DTC states are degenerate, this choice of label is completely arbitrary. We introduce the notation $[I_1, O, I_2]$ to denote the state of the three-site DLM at a given time. $I_{1, 2} \in \{ 0, 1 \}$ refers to the state of the input sites, as shown in Fig. \ref{fig:fig6}(a), while $O \in \{ 0, 1 \}$ refers to the state of the output site. For a NAND (NOR) operation, the output site $O$ is initialized as a 1-bit (0-bit), while the input sites $I_1$ and $I_2$ are initialized according to the desired computation. 

The exemplary dynamics in Fig.~\ref{fig:fig6}(b) show the mean-field dynamics associated with the NAND gate operation for the input [111]. As soon as the $\phi_{\ell}$ of all sites have reached a steady state, we perform a logic operation by first applying a long pulse on the coupling between site $I_{1}$ and $O$ with strength $J_{1O}$. Then, immediately after the long pulse, we couple site $I_{2}$ to $O$ with coupling strength $J_{2O}$ for a shorter duration $T_{2O}$. The long pulse strength is chosen such that $J_{c,\mathrm{FM}} < J_{1O} < J_{c,\mathrm{AFM}}$, while its duration, $T_{1O}$, is set to exceed $\tau_{r,\mathrm{AFM/FM}}$. This ensures that $I_1$ and $O$ will have relaxed to an NP by the start of the second, shorter pulse if $I_1 = 1$ and are thus initially in the FM-DTC state. In this case, the short pulse then serves to anti-synchronize $O$ with $I_2$ for a suitable choice of $T_{2O}$. This yields an output state of $O=0$ ($O=1$) for an input of $I_{2} = 1$ ($I_{2} = 0$), which is consistent with the required NAND output for the initial states [111] and [110], as listed in Fig.~\ref{fig:fig6}(a). If $I_{1} = 0$ instead, then the output site remains in the DTC state as a 1-bit during the long pulse since $J_{1O} < J_{c, \mathrm{AFM}}$, and thus will not be affected by the second pulse provided that $J_{2O}$ is sufficiently small. This recreates the correct NAND output for the remaining cases [011] and [010]. $O$ can then be read once it has reached a steady state after the second pulse, yielding the final output. These results are consistent with the truth table shown in Fig.~\ref{fig:fig6}(a), which indicates that the output site must only be flipped to $O = 0$ for [111] in order to realize a NAND gate. We also note that applying the transformation $0 \rightarrow 1$ to the NAND truth table yields the NOR algebra. Such a transformation is always valid for our system since the two DTC states are degenerate. This ambiguity can be resolved by introducing an additional reference node with a reference state chosen such that the gate emulates either a NAND or NOR gate \cite{emman_2026_singlepulse}. Following this, throughout the remaining section, we will solely focus on the NAND case since the same arguments applies for the NOR case.

In Fig.~\ref{fig:fig6}(c), we demonstrate how we can determine the optimal $J_{1O}$ for the long pulse by examining the behavior of $A_r$ and $\tau_r$. For a successful NAND operation, we see that the coupling between $I_{1}$ and $O$ must be $J_{c, \mathrm{FM}} < J_{1O} < J_{c, \mathrm{AFM}}$. For the chosen spin-boson coupling, $\lambda_{0} = 0.9\lambda_{c}$, we find that there is no $J_{c, \mathrm{AFM}}$ within the range $0 < J < 0.5\omega$. We do not consider the region $J > 0.5\omega$ in line with the results in Ref.~\cite{zou_2014_dlm}, in which above $J> 0.5\omega$, the NP becomes unstable, and thus the Holstein-Primakoff approximation used to obtain Eq.~(3) fails. In order to minimize $T_{1O}$, we choose $J_{1O}$ so that the relaxation times for [01X] and [11X] are at a minimum, which we find to be at around $J_{1O} = 0.45\omega$ for the given set of parameters in Fig.~\ref{fig:fig6}(c). Here, the `X' symbol indicates that the state of $I_2$ is irrelevant since we are currently interested in the dynamics during the long pulse. Then, revisiting Fig.~\ref{fig:fig5}(a), we can set $T_{2O}$ such that $\Delta \phi = \pi$.

In order to test the robustness of the protocol in the presence of quantum fluctuations, we consider the finite-size limit using the TWA. We present in Fig.~\ref{fig:fig6}(d) an exemplary single-trajectory TWA dynamics for the same parameters and initial states as Fig.~\ref{fig:fig6}(b). Notably, the introduction of noise can cause the protocol to fail for the case in which the initial DTC states are [111]. This failure stems from the system's behavior during the long pulse, in which the FM-DTC [11X] acquires a finite lifetime $\tau$ due to the fluctuations pushing it towards the stable [01X] or [10X] state instead of the intended NP. This behavior is unwanted, as the protocol can no longer distinguish between the inputs [01X] and [11X] since memory of the original state is lost during the long pulse. To minimize the likelihood of this failure, $T_{1O}$ must be set to $T_{1O} < \tau$ for the system to remain in the [11X] state following the long pulse. In addition, we want $\tau_{r, \mathrm{FM}} \ll T_{1O}$ so that the amplitude of $O$ reduces quickly. This can be accomplished by choosing $J_{1O} \gg J_{c, \mathrm{FM}}$ since $\tau_{r, 11}$ diverges near the critical point.

\begin{figure}\label{fig:Dicke_Validity}
    \centering
    \includegraphics[scale = 0.29]{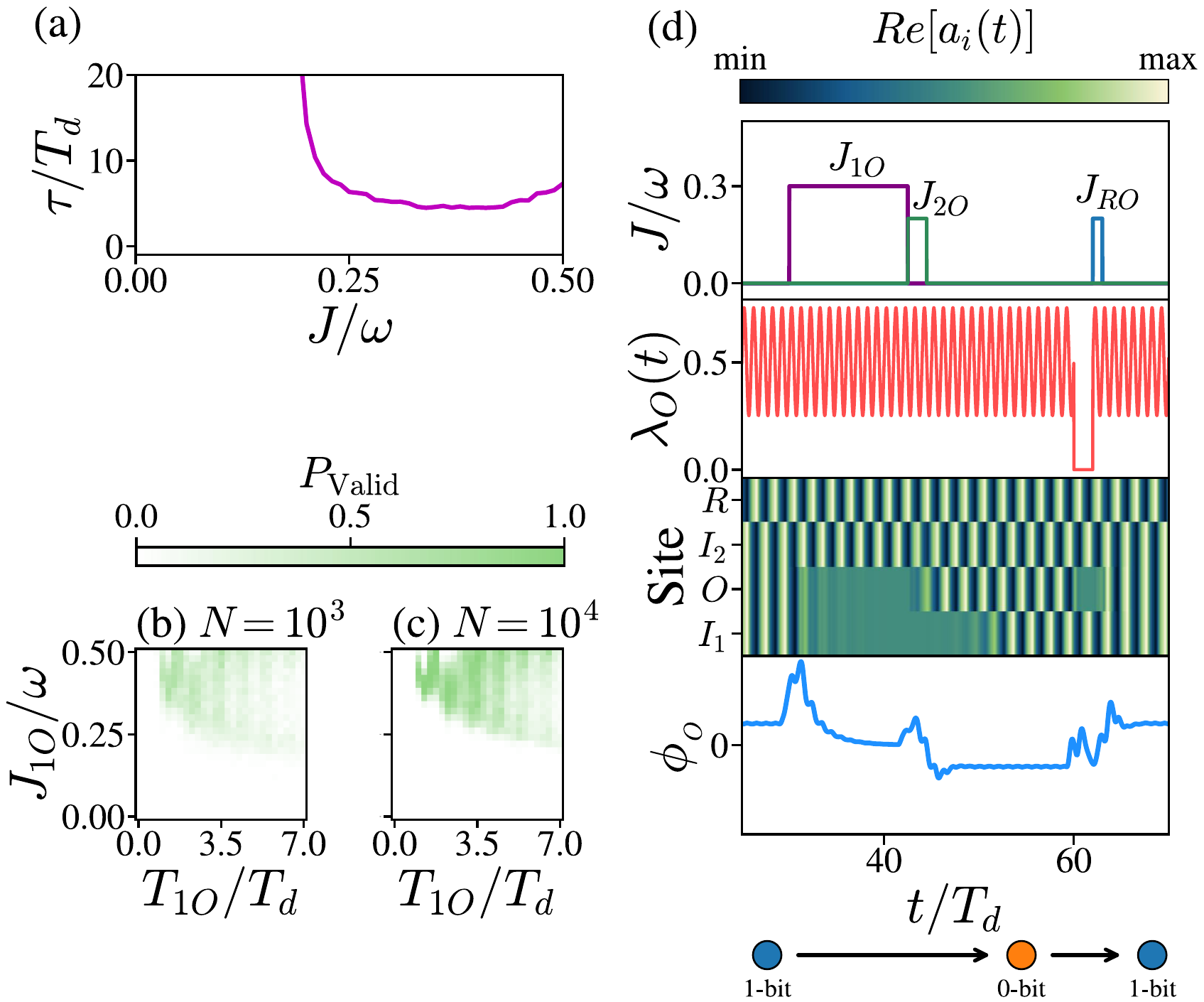}
    \caption{(a) $\tau$ for the FM-DTC as a function of $J$ when $N =10^5$. (b)-(c) The probability $P_{\mathrm{Valid}}$ of obtaining a valid logic gate calculation as a function of $J_{1O}$ and $T_{1O}$ when (b) $N = 10^3$ or (c) $N = 10^4$. (d) Exemplary dynamics showing the NAND calculation for the input configuration [111] and how the reset protocol can be used to reset the output site to the $1$-bit state afterwards. (Top panel) The coupling $J_{1O}$, $J_{2O}$, and $J_{RO}$. (Middle panels) The driving strength for the output site $\lambda_{O}(t)$ and the dynamics of $Re[a_i(t)]$ for each site. (Bottom panel) The absolute time phase $\phi_O$ for the output site. The state of the output site is represented through the colored balls underneath the bottom panel. The remaining parameters are $\{\omega_0, \omega_d, A, N, \lambda_0\} = \{\omega, 0.8\omega, 0.5, 10^5, 0.9\lambda_c\} $.}
    \label{fig:fig7}
\end{figure}

To identify the $T_{1O}$ at which the logic operation remains robust for the parameters in Fig.~\ref{fig:fig6}(c), we obtain in Fig.~\ref{fig:fig7}(a) the behavior of $\tau$ for varying $J$. To preserve the validity of the logic gate protocol in the presence of noise, $J_{1O}$ must be optimized such that the amplitude is minimized during the coupling, while $\tau$ is maximized. These conditions are satisfied when $J_{1O} \gg J_{c, \mathrm{FM}}$. In Figs.~\ref{fig:fig7}(b)-(c), we show the probability of obtaining a valid NAND gate $P_\mathrm{Valid}$ as a function of $T_{1O}$ and $J_{1O}$ using the same realization of noise for all four input configurations. As mentioned earlier, we see that if we increase $T_{1O}$ past $\tau$, the probability of success drops drastically since the quantum fluctuations can flip $O$ for the input [011]. Below $J_{1O} = J_{c, \mathrm{FM}}$, we have no successful NAND operation since the consecutive pulsed coupling between $I_{1, 2}$ and $O$ cannot convert the DTC in the output site into a NP for [111], and thus cannot flip the state of the site as required for a NAND gate. Increasing $N$ from $N = 10^3$ to $N = 10^4$ increases the success probability, since the relative strength of the quantum fluctuations decreases with increasing $N$. Furthermore, we find regions in the parameter space in which the protocol exhibits robustness against noise leading to $100\%$ success rate for the NAND gate.

Finally, to perform a subsequent calculation, the input and output sites are re-initialized accordingly. This can be accomplished by performing the protocol shown in Fig.~\ref{fig:fig7}(d). Here, the  periodic driving in the output site $\lambda_O(t)$ is removed, thereby melting the DTC in the site. Then, by performing the anti-synchronization or synchronization protocol outlined in Fig.~\ref{fig:fig1}(e) between the reference and output sites, the output site can be initialized as either a 0-bit or 1-bit by choosing an appropriate value for the coupling time $T_{RO}$ between the output site and the reference site. The input sites $I_1$ and $I_2$ can be re-initialized using the same procedure.

\section{Conclusion}\label{sec:sec6}

We have explored the dynamics of coupled DTCs in the two-site DLM. We have shown that coupling together two DTCs can decrease their oscillation amplitude depending on the system's parameters and the initial configuration. We have examined the relaxation timescale of the oscillation amplitude of the two coupled DTCs and found divergences near the critical coupling strengths $J_{c, \mathrm{FM/AFM}}$ above which the DTCs become unstable towards the formation of an NP, implying a continuous phase transition. We have then shown that a site initialized in the NP can be made to synchronize or anti-synchronize with a DTC by applying a pulsed coupling, in which the two sites are connected for a finite duration. Following these results, we have proposed a protocol to implement a classical NAND and NOR gate architecture using three DTCs by performing consecutive pulsed coupling between each of the input sites and the output site. We have demonstrated its robustness against quantum fluctuations in a wide region of parameter space, thus showing the potential of using coupled DTCs in complex logic gate operations.

\textit{Acknowledgement. - } This work was primarily funded by the Department of Science and Technology (DOST) and monitored by the Philippine Council for Industry, Energy, and Emerging Technology Research and Development (DOST-PCIEERD) with Project No. 1214356.

\bibliography{reference.bib}
\bibliographystyle{apsrev4-2}

\end{document}